\documentclass[conference]{IEEEtran}
\IEEEoverridecommandlockouts
\usepackage{cite}
\usepackage{amsmath,amssymb,amsfonts}
\usepackage{algorithmic}
\usepackage{graphicx}
\usepackage{textcomp}
\usepackage{xcolor}
\usepackage[ruled,linesnumbered]{algorithm2e}
\usepackage{graphicx}
\usepackage{booktabs}
\usepackage{colortbl}
\usepackage{diagbox}
\usepackage{array}
\usepackage{color}
\usepackage{graphicx}
\usepackage{multirow}
\usepackage{ragged2e}
\usepackage{amsthm}
\usepackage{hyperref}
\usepackage{xcolor}
\usepackage{tabu}

\theoremstyle{definition}
\newtheorem{definition}{Definition}[section]

\graphicspath{{figures/}}

\newcommand{\ourmethod}{\texttt{MAMDR}\xspace}
\def\BibTeX{{\rm B\kern-.05em{\sc i\kern-.025em b}\kern-.08em
    T\kern-.1667em\lower.7ex\hbox{E}\kern-.125emX}}
\begin{document}

\title{MAMDR: A Model Agnostic Learning Framework for Multi-Domain Recommendation}

\author{\IEEEauthorblockN{Linhao Luo$^{1*}$ \thanks{*~This work was conducted during an internship in Alibaba Group.}, Yumeng Li$^2$, Buyu Gao$^2$, Shuai Tang$^3$, Sinan Wang$^2$, Jiancheng Li$^2$, Tanchao Zhu$^2$, Jiancai Liu$^2$,\\ Zhao Li$^{4\dag}$\thanks{\dag~Corresponding authors}, and Shirui Pan$^{5\dag}$}\\
\IEEEauthorblockA{\textit{$^1$Monash University, $^2$Alibaba Group, $^3$Independent Researcher}
\textit{$^4$Hangzhou Link2Do Technology, $^5$Griffith University}\\
linhao.luo@monash.edu, bellium@163.com, lzjoey@gmail.com, s.pan@griffith.edu.au\\ 
\{lym174806,gaobuyu.gby,sinan.wsn,ljc250108,tanchao.zhutc,jiancai.ljc\}@alibaba-inc.com
}
}


\maketitle

\begin{abstract}
    Large-scale e-commercial platforms in the real-world usually contain various recommendation scenarios (domains) to meet demands of diverse customer groups. Multi-Domain Recommendation (MDR), which aims to jointly improve recommendations on all domains and easily scales to thousands of domains, has attracted increasing attention from practitioners and researchers. Existing MDR methods usually employ a shared structure and several specific components to respectively leverage reusable features and domain-specific information. However, data distribution differs across domains, making it challenging to develop a general model that can be applied to all circumstances. Additionally, during training, shared parameters often suffer from domain conflict while specific parameters are inclined to overfitting on data sparsity domains.
    In this paper, we first present a scalable MDR platform served in Taobao that enables to provide services for thousands of domains without specialists involved.
    To address the problems of MDR methods, we propose a novel model agnostic learning framework, namely \ourmethod, for the multi-domain recommendation. Specifically, we first propose a Domain Negotiation (DN) strategy to alleviate the conflict between domains. Then, we develop a Domain Regularization (DR) to improve the generalizability of specific parameters by learning from other domains. We integrate these components into a unified framework and present \ourmethod, which can be applied to any model structure to perform multi-domain recommendation. Finally, we present a large-scale implementation of \ourmethod in the Taobao application and construct various public MDR benchmark datasets which can be used for following studies. Extensive experiments on both benchmark datasets and industry datasets demonstrate the effectiveness and generalizability of \ourmethod.
\end{abstract}


\section{Introduction}\label{sec:introduction}
Recommender systems have been widely applied in many applications to provide personalized experiences for users. Conventional recommender systems \cite{cheng2016wide,he2017neural,xin2021atnn} are trained and predicted on samples collected from a single domain.
However, large e-commercial platforms such as Taobao and Amazon need to build recommender systems for various domains to satisfy diverse user demands and stimulate users' purchases. For example, Taobao contains multiple business domains such as ``what to take when traveling'', ``how to dress up yourself for a party'', and ``things to prepare when a baby is coming''. Each domain contains a set of products that are related to the domain's topic and promotion strategies to stimulate purchasing.
Thus, multiple domains are expected to be jointly considered to enable effective recommendation. Unlike cross-domain recommendation (CDR) \cite{hu2018conet} that only focuses on one or more target domains,  multi-domain recommendation (MDR) \cite{tang2011analysis,sheng2021one} aims to simultaneously improve recommendation results of all domains.

\begin{figure}
    \centering
    \includegraphics[trim=0 0.3cm 0.8cm 0.2cm,clip,width=0.9\columnwidth]{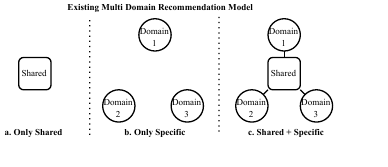}
    \caption{Existing Muti-Domain recommender models. (a) one model for all domains, (b) one model for each domain, (c) one model with shared and domain-specific parameters.}
    \label{fig:mdr_model}
    \end{figure}

The main challenge of MDR is that the data distribution is not consistent across domains. Because distinct domains have only partially overlapping user and item groups, and varied domain marketing tactics result in diverse user behavior patterns. For example, during the domain of ``Singles Day Promotion (Double 11)'', China's largest online shopping event, customers are more inclined to click recommended goods due to the significant discounts, whereas some minor domains could have fewer user activities. These distinctions result in domain-specific data distributions and pose challenges for using the same model for all domains, as shown in Figure \ref{fig:mdr_model} (a). A simple solution is to build a separate model for each domain as shown in Figure \ref{fig:mdr_model} (b). However, some domains do not have enough data to optimize a great separated model \cite{xu2022metacar}; meanwhile, the separated models disregard the shared information between domains and lead to a sub-optimal result \cite{wang2022multi}. Moreover, designing a model for each domain causes tremendous resource consumption for model storage and requires specialists to find the optimal model for each domain, which is very time-consuming and unscalable. Thus, MDR methods, which could easily scale to thousands of domains without human effort, have attracted increasing attention.

Conventional MDR methods \cite{zhang2010multi,ma2018modeling,sheng2021one,wang2022multi}, inspired by Multi-task learning (MTL), treat each domain as a task. Most of them split model parameters into a set of shared parameters and domain-specific parameters, as shown in Figure \ref{fig:mdr_model} (c). The shared parameters are optimized by using all domains' data to leverage multi-domain features, and the specific parameters are optimized by using domain-specific data to capture the domain distinction. In this way, we can support multiple domains by adding specific parameters. However, these methods still have the following limitations: 
\begin{enumerate}
    \item Existing MDR models cannot generalize to all circumstances.
    \item Shared parameters suffer from the domain conflict, and specific parameters are inclined to overfitting.
    \item Lack of public MDR benchmark datasets.
\end{enumerate}
\textbf{(Limit. 1)} previous research \cite{misra2016cross} shows that the structure of shared and specific parameters are diverse in domains and essential to final results. Due to the distinctive data distributions, it is difficult for existing MDR models to accommodate all circumstances.
\textbf{(Limit. 2)} Shared parameters suffer from the domain conflict problem. The gradients from each domain could point to dissimilar directions. This ends up causing the shared parameters to stay at a compromised position on the optimization landscape and deteriorate the overall performance. The specific parameters are separately optimized on each domain's data. Thus, they can easily overfit on data sparsity domains and cannot perform well to unseen data.  \textbf{(Limit. 3)} important as the MDR problem could be, very few public datasets exist. Most existing methods are evaluated on private industry datasets \cite{sheng2021one}, which is difficult for following researchers to compare with.

Aforementioned problems limit the application of MDR methods in industries and other data engineering areas. In this paper, we first present the architecture of the multi-domain recommender system in Taobao. This system is served as a scalable MDR platform that enables to provide services for thousands of domains without specialists involved. Then, to address the limitations of existing MDR methods, we propose a novel model agnostic learning framework for the multi-domain recommendation, denoted as \ourmethod. \ourmethod does not require any constraints on the model structure, it can be readily applied to any existing recommender model which makes it a perfect fit for the MDR system in the industry (to address \textbf{Limit. 1}). Specifically, we simultaneously consider the optimization of shared parameters and specific parameters (to address \textbf{Limit. 2}). We first propose the Domain Negotiation (DN), which mitigates domain conflict by effectively maximizing the inner-products of gradients between domains. Second, we present a strategy denoted as Domain Regularization (DR) for specific parameters. DR enables the optimization process on other domains to minimize the loss on the specific domain. Thus, DR can alleviate the overfitting problem on data sparsity domains. The effectiveness of DN and DR is proved by both theoretical analyses as well as empirical studies. 

To support large-scale applications, we integrate DN and DR into a unified framework (\ourmethod) and provide a disturbed implementation of \ourmethod. Especially, we introduce the dynamic-cache and static-cache mechanisms to reduce the synchronization overhead and alleviate inconsistency, which would facilitate the training in industry scenarios. 
Finally, to facilitate the research in MDR and other relative data engineering tasks, we present various MDR benchmark datasets (to address \textbf{Limit. 3}). These datasets simulate real-world situations based on the challenges we found in our industrial applications, which could help future researchers.

The main contributions of this paper are summarized as follows:
\begin{itemize}
    \item We present a multi-domain recommender system served in Taobao, and propose a novel model agnostic learning framework: \ourmethod, which is compatible with arbitrary model structures. A distributed implementation of \ourmethod is also provided to support large-scale applications.
    \item We propose two scalable algorithms: Domain Negotiation (DN) and Domain Regularization (DR) to alleviate the domain conflict and overfitting problem in MDR. Theoretical analyses are also provided to ensure effectiveness.
    \item We have provided various benchmark datasets to simulate the real-world challenges in MDR problems. Extensive experiments on public datasets and a large-scale industry dataset demonstrate both the effectiveness and scalability of \ourmethod.
\end{itemize}
\section{Related Work}\label{sec:relatedwork}

\subsection{Multi-Domain Recommendation}
Recommender system has been a long-standing research topic \cite{luo2020motif,yao2020correlated,li2021large,yang2022semantically,luo2022dcrs,li2020hierarchical}. Recently, the MDR problem has garnered considerable attention. Previous methods \cite{zhang2010multi, ma2018your, xie2020internal} either focus on investigating the domain connections, or try to leverage useful features between domains \cite{hao2021adversarial}.
Additionally, by considering each domain as a task, multi-task approaches (e.g., Shared-Bottom \cite{ruder2017overview}, MMoE \cite{ma2018modeling}, and PLE \cite{tang2020progressive}) could be simply deployed to tackle the MDR problem. Inspired by MTL, STAR \cite{sheng2021one} separates the model parameters into shared and domain-specific parts. But it still suffers from domain conflict and overfitting problems. CMoIE \cite{wang2022multi} extends the framework of MMoE with conflict resolution modules, which requires modifying the model structure.
Additionally, they are evaluated on private industry datasets.
Related to MDR, cross-domain recommendation (CDR) aims to improve the performance of target domains with the help of auxiliary domains \cite{khan2017cross,hu2018conet,du2019sequential}. By treating each domain as the target domain, CDR methods can be adapted to the MDR problem. However, the time complexity of applying the CDR method to address the MDR problem is unacceptable.

\begin{figure*}[]
    \centering
    \includegraphics[trim=0cm 0.0cm 0.0cm 0,clip,width=0.9\linewidth]{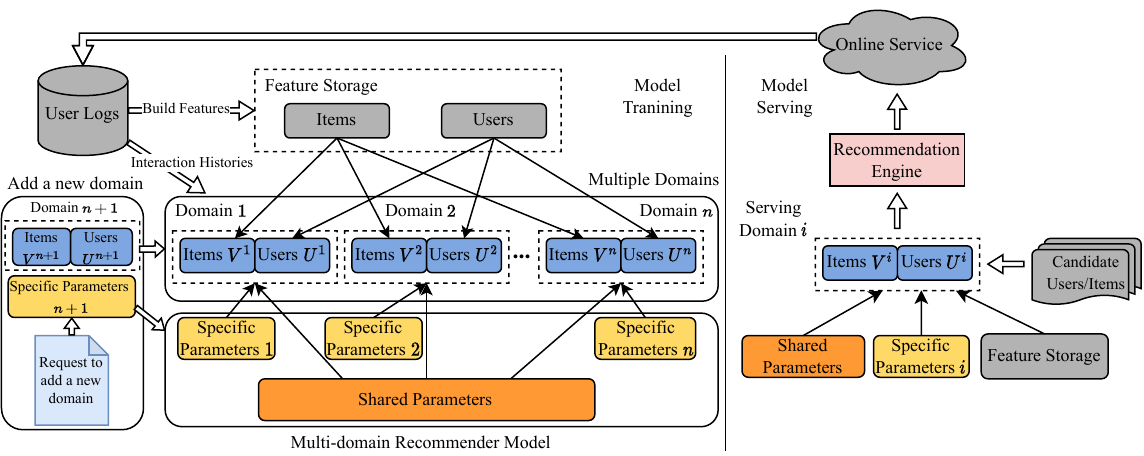}
    \caption{The framework of Multi-Domain Recommender system served in Taobao.}
    \label{fig:framework}
\end{figure*}

\subsection{Multi-Domain Learning}

Multi-Domain Learning (MDL) \cite{yang2015unified} has been widely applied in the real-world. Some MDL research can be extended to solve the problems in MDR. 
The domain generalization (DG) methods seek to distill the common knowledge from multi-domains and learn more robust features that are potentially useful for unseen domains. Existing research \cite{li2017deeper} assumes that any domain can be factorized into a shared and domain-specific component. Mansilla et al. \cite{mansilla2021domain} analyze the multi-domain gradient conflict and adapt the PCGrad \cite{yu2020gradient} into the multi-domain setting. The PCGrad alleviates gradient conflicts of two domains by projecting their gradients into the non-conflict direction. However, this strategy is inefficient for scenarios with more than two domains. MLDG \cite{li2018learning} improves the generalizability of learned parameters by extending the idea of meta-learning. By using meta-learning, Dou et al. \cite{dou2019domain} introduce two complementary losses to explicitly regularize the semantic structure of the feature space.  Similarity, Sicilia et al. \cite{sicilia2021multi} adopt meta-learning to balance losses between different domains. MetaReg \cite{balaji2018metareg} also adopts the meta-learning as a regularization term to achieve good multi-domain generalization.

\subsection{Meta-Learning}

Meta-learning methods (e.g., MAML \cite{finn2017model} and Reptile \cite{nichol2018first}) aim to learn generalized initialized weights that can be readily tailored to new tasks, which is agnostic to model structure. Meta-learning acquires the shared knowledge across tasks and enables specific information through a few-step finetuning \cite{sun2019meta}. Meta-learning may be used to the MDR problem by treating each task as a domain. To address the gradient conflict, MT-net \cite{lee2018gradient} enables the parameters to be learned on task-specific subspaces with distinct gradient descent. WarpGrad \cite{flennerhag2019meta} further proposes a Warped Gradient Descent that facilitates the gradient descent across the task distribution. L2F \cite{baik2020learning} introduces a task-and-layer-wise attenuation mask on the initial parameters to alleviate the conflict among tasks. GradDrop \cite{tseng2020regularizing} presents a meta-learning-based Gradient Dropout to avoid overfitting for certain tasks. TADAM \cite{oreshkin2018tadam} develops a metric scaling method to provide task-dependent metric space for optimizing specific parameters. HSML \cite{yao2019hierarchically} introduces a hierarchical task clustering structure to preserve generalization knowledge among tasks, while also maintaining the specific information.

Even some multi-domain learning and meta-learning frameworks can be applied to the MDR problem, they are not as effective as the proposed \ourmethod. Our method enables scalable implementation in the large-scale MDR problem and is compatible with the existing recommender models.

\section{Preliminary}

\subsection{Multi-Domain Recommendation and Applications}\label{sec:app}
Multi-Domain Recommendation (MDR) \cite{zhu2021cross} aims to design a system that recommends a group of items to a set of users from multiple domains. The recommender system satisfies diverse user demands and provides personalized experiences under each domain, meanwhile, it can be easily scaled to new domains. 

The Multi-Domain Recommender system (MDR system) served in Taobao is illustrated in Figure \ref{fig:framework}. In our applications, we need to provide services for thousands of different domains, where some of which have limited training data. In training, the recommender model is optimized using user-item interaction histories parsed from the user logs. The interaction histories are collected from different domains. Different domains could share overlapping users/items. Thus, we maintain a global feature storage for all users/items and shared model parameters to server for all domains. We also design specific parameters to provide customized recommendations under each domain. A new domain can be easily added to the system by providing the corresponding users/items. The system would automatically increase specific parameters for this new domain.
However, how to incorporate the shared and specific components together while optimizing them under all domains remains a significant challenge in Multi-Domain Recommendation. The problem of Multi-Domain Recommendation can be defined as:

\begin{definition}[\textbf{Multi-Domain Recommendation}]
Given $n$ different domains $\mathcal{D}=\{D^1,\cdots,D^n\}$, each domain $D^i = \{U^i, V^i, T^i\}$ includes a set of users $u \in U^i$ and items $v \in V^i$, where users and items could overlap across domains. The $T^i$ denotes a set of user-item interaction histories $(u,v,y)\in T^i$, where $y\in\{1,0\}$ indicates whether user $u$ clicked item $v$ or not. Thus, Multi-Domain Recommendation aims to use interaction histories of all domains $\mathcal{T}=\{T^1,\cdots T^n\}$ to train a model with parameter $\Theta$ that could recommend interesting items for users in all domains. 

As we discussed in the section \ref{sec:introduction}, some MDR methods split the model parameters $\Theta$ into a set of shared parameters $\theta^S$ and a set of domain-specific parameters $\{\theta^i| i \in [1,n]\}$. The $\theta^S$ are optimized by data from all domains to capture multi-domain features, and the $\theta^i$ is updated on specific domain to capture distinction. Thus, the objective function of MDR can be further formulated as:
    \begin{equation}
        \mathcal{O}_M = \min_{\Theta=\{\theta^S,\theta^i\}} \sum_{i=1}^n L(\theta^S,T^i) + L(\theta^i,T^i),
        \label{eq:objective_specific}
    \end{equation}
    where $\theta^S$ are optimized by data from all domains, and $\theta_i$ are only optimized in corresponding domain. 
\end{definition}

\subsection{Domain Conflict}\label{sec:conflict}

\begin{figure}[]
    \centering
    \includegraphics[trim=0 0.5cm 0.8cm 0cm,clip,width=0.5\columnwidth]{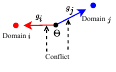}
    \caption{The illustration of domain conflict. The conflict happens when the inner-product of gradients from different domains is negative.}
    \label{fig:domain_conflict}
\end{figure}

Directly optimizing equations \ref{eq:objective_specific} may deteriorate the recommendation performance. A primary cause for this phenomena is known as domain conflict, which is shown in the Figure \ref{fig:domain_conflict}.
For parameters optimized across domains, the gradient from each domain $D^i$ is denoted as $g_i = \bigtriangledown L(\Theta,T^i)$. The $\Theta$ are optimized following the direction of $g_i$, i.e., $\Theta \leftarrow  \Theta - \alpha \cdot g_i$, where $\alpha$ is the learning rate. However, the gradients from different domains may conflict with each other when they point to dissimilar directions. This dissimilarity could be represented by the inner-product between gradients. Thus, the conflict happens when the inner-product of gradients from different domains is negative, i.e., $\langle g_i, g_j\rangle < 0$, where $\langle\cdot,\cdot\rangle$ denotes the inner-product between two gradients. As observed by previous research \cite{yu2020gradient,liu2021conflict}, this conflict will impair the optimization and lead parameters to stay at a compromise point at the loss landscape, which also results in poor recommendation results.

\subsection{Prior Attempts and Limitations}\label{sec:limit}
Some research efforts have been made to solve the domain conflict. In the area of MTL, Alex et al. \cite{kendall2018multi} utilize a learned weighted loss to balance the gradients from different domains. PCGrad \cite{yu2020gradient} relieves the conflict by projecting the gradient into the normal plane of others. In the area of meta-learning, the conflicts of gradients could be averse by projecting them into a common space \cite{flennerhag2019meta} or minimizing the inner-product of them \cite{li2018learning}.

However, MTL methods that manipulate gradients could face the convergence problem and stay at a sub-optimal point of the loss landscape \cite{liu2021conflict}. Meanwhile, meta-learning frameworks are originally proposed to apply in unseen domains, which might not fit existing domains' data well. Besides, the above methods are either required to modify the model structure or lack scalability for large MDR. Despite some MDR methods using domain-specific parameters, their share parameters $\theta^S$ still suffer the aforementioned problems. Besides, the uses of specific parameters also meet the overfitting problem when the domain data is insufficient. 

Related to MDR, cross-domain recommendation (CDR) aims to improve the performance of the target domain with the help of auxiliary domains \cite{khan2017cross,loni2014cross,hu2018conet,du2019sequential,liu2021leveraging}. By treating each domain as the target domain and transferring knowledge from each auxiliary domain, CDR methods can be adapted to MDR problems. However, it requires a $O(n^2)$ complexity, which is unacceptable for large-scale applications. We can transfer from multiple domains at a time to reduce complexity, but it also introduces the domain conflict problem.

Thus, in \ourmethod, we introduce the Domain Negotiation (DN) and Domain Regularization (DR) strategies to solve the aforementioned challenges in a linear time complexity.

\section{Approach}\label{sec:approach}
In this section, we will first introduce the Domain Negotiation and Domain Regularization in subsections \ref{sec:dn} and \ref{sec:dr}, respectively. Then, the theoretical analyses for DN and DR will be discussed in subsection \ref{sec:dn_analysis}. Last, we will introduce the overall algorithm of \ourmethod and the large-scale implementation of \ourmethod in subsection \ref{sec:mamdr}. 

\subsection{Domain Negotiation (DN)}\label{sec:dn}
Domain Negotiation (DN) is proposed to mitigate the domain conflict problem. Given $n$ different domains, the Domain Negotiation (DN) is performed as shown in the Algorithm \ref{alg:dn}.

\begin{algorithm}[]
    \caption{Domain Negotiation (DN)}\label{alg:dn}
    \KwIn {$n$ different domains $\mathcal{D}$, initial model parameters $\Theta$, learning rate $\alpha$ and $\beta$, maximum training epoch $N$.}
    \KwOut{$\Theta$}
    \For{$epoch = 1,\cdots, N$}{
        $\widetilde{\Theta}_1 \gets \Theta$\;
        Randomly shuffle $\mathcal{D}$\;
        \For{$i\leftarrow 1,\cdots,n$}{\label{alg:inner}
            Update $\widetilde{\Theta}_{i+1} \gets \widetilde{\Theta}_i - \alpha \cdot \bigtriangledown L(\widetilde{\Theta}_i, T^i)$\;\label{eq:inner}
        }
        Update $\Theta \gets \Theta + \beta \cdot (\widetilde{\Theta}_{n+1} - \Theta)$\;
    }
    \Return{$\Theta$}
\end{algorithm}

As shown in the Algorithm \ref{alg:dn}, DN consists of two training loops: the \textit{outer loop} (line 1-8) and \textit{inner loop} (line 4-6). At the beginning of each inner loop, the $\widetilde{{\Theta}}_1$ are initialized by ${\Theta}$. Then, during the inner loop, the $\widetilde{\Theta}_i$ are sequentially updated on each domain $i$ with random order, which can be formulated as:
\begin{equation}
    \widetilde{\Theta}_{i+1} \gets \widetilde{\Theta}_i - \alpha \cdot \bigtriangledown L(\widetilde{\Theta}_i, T^i),
\end{equation}
where $T^i$ is the data from domain $i$, and $\alpha$ denotes the inner-loop learning rate.
After the inner loop, we treat $\widetilde{\Theta}_{n+1} - \Theta$ as the gradient for outer loop optimization, which directly updates the parameters $\Theta$. This can be formulated as:
\begin{equation}
    \Theta \gets \Theta + \beta \cdot (\widetilde{\Theta}_{n+1} - \Theta),\label{eq:outer}
\end{equation}
where $\beta$ denotes the outer-loop learning rate. Both the inner loop and outer loop can use arbitrary optimizers, such as traditional SGD, Adam or Parallelized SGD \cite{bottou2010large} for distributed training in large-scale applications.

Noticeably, when $\beta$ is set to 1, DN will degrade to Alternate Training (one-by-one training) \cite{long2021multi}, which could corrupt the performance. We discuss the reason and show the empirical results in subsections \ref{sec:dn_analysis} and \ref{sec:parameters}, respectively.

\subsection{Domain Regularization (DR)}\label{sec:dr}
Despite DN being a powerful strategy for mitigating domain conflict of shared parameters, the specific parameters are still prone to overfit on data sparsity domains.
In this section, we will introduce Domain Regularization (DR) for optimizing domain-specific parameters, which greatly improves the performance of \ourmethod.

\begin{figure}[h]
    \centering
    \includegraphics[trim=0cm 0.5cm 1.2cm 0,clip,width=1\columnwidth]{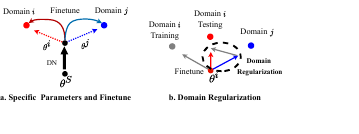}
    \caption{(a) The similarity between domain-specific parameters and finetune. (b) The proposed Domain Regularization.}
    \label{fig:dr}
\end{figure}

Traditionally, after optimizing across domains, the model will be finetuned on each specific domain to obtain several domain-specific models. Recently, some MDR methods \cite{sheng2021one} proposed to use domain-specific parameters to replace the finetune process. The domain-specific parameters $\theta^i$ have the same dimension as the shared parameters $\theta^S$, but they are only optimized by domain-specific data. The $\theta^i$ are integrated with shared parameters $\theta^S$ by an element-wise add operation, which can be formulated as:
\begin{equation}
    \Theta = \theta^S + \theta^i.
\end{equation}
As shown in Figure \ref{fig:dr} (a), the $\theta^i$ can be treated as a direction pointing to the endpoint of the finetune process, thus it can achieve similar results as finetune.

However, one major issue of finetuning is that it is easy to overfit on some data sparsity domains. For example, as shown in Figure \ref{fig:dr} (b), the $\theta^i$ is optimized on a data sparsity domain $i$. Though $\theta^i$ could perfectly fit on the training data of domain $i$, it cannot generalize well on the testing data as shown in the gray arrow. The Domain regularization (DR), as circled in Figure \ref{fig:dr} (b), optimizes $\theta^i$ with the help of other domains' data to improve $\theta^i$'s generalizability. The details of Domain Regularization are shown in Algorithm \ref{alg:dr}.

Given a target domain $D^i$ and its specific parameters $\theta^i$, DR samples $k$ different domains $\widetilde{\mathcal{D}}$ from $\mathcal{D}$. For each $D^j \in \widetilde{\mathcal{D}}$, DR first update $\widetilde{\theta^i}$ on $D^j$, then update it on $D^i$ as regularization, which can be formulated as:
\begin{gather}
    \widetilde{\theta^i} \gets \theta^i,\\
    \widetilde{\theta^i} \gets \widetilde{\theta}^i - \alpha \cdot \bigtriangledown L(\widetilde{\theta}^i, T^j),\\
    \widetilde{\theta^i} \gets \widetilde{\theta}^i - \alpha \cdot \bigtriangledown L(\widetilde{\theta}^i, T^i).
\end{gather}
At last, the $\widetilde{\theta^i} - \theta^i$ is denoted as the gradient to update $\theta^i$, which can be formulated as:
\begin{equation}
    \theta^i \gets \theta^i + \gamma \cdot (\widetilde{\theta^i} - \theta^i),
\end{equation}
where $\gamma$ is the learning rate for DR. 

Unlike DN, in which the domain order is random at each iteration, the optimization sequence is fixed in DR. We first update specific parameters on domain $j$, then update them on the target domain $i$. In this way, we can make sure that only the helpful information from domain $j$ is extracted for the target domain. The detailed analysis can be found at section \ref{sec:dn_analysis}.

\begin{algorithm}[h]
    \caption{Domain Regularization (DR)}\label{alg:dr}
    \KwIn {$n$ different domains $\mathcal{D}$, target domain $D^i$, specific parameters $\theta^i$, learning rate $\alpha,\gamma$, sample number $k$}
    \KwOut{$\theta^i$}
    Sample $k$ domains from $\mathcal{D}$ as $\widetilde{\mathcal{D}}$\;
    \For{$D^j$ in $\widetilde{\mathcal{D}}$}{
        $\widetilde{\theta^i} \gets \theta^i$\;
        Update $\widetilde{\theta^i} \gets \widetilde{\theta}^i - \alpha \cdot \bigtriangledown L(\widetilde{\theta}^i, T^j)$ \# Update on domain $j$\;
        Update $\widetilde{\theta^i} \gets \widetilde{\theta}^i - \alpha \cdot \bigtriangledown L(\widetilde{\theta}^i, T^i)$ \# Using domain $i$ as regularization\;
        Update $\theta^i \gets \theta^i + \gamma \cdot (\widetilde{\theta^i} - \theta^i)$\;
    }
    \Return{$\theta^i$}
\end{algorithm}

\begin{figure*}[]
    \centering
    \includegraphics[trim=0.2cm 0.0cm 0.0cm 0,clip,width=0.75\linewidth]{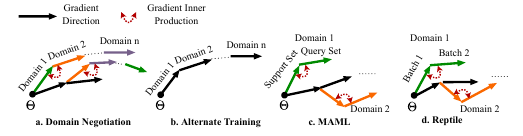}
    \caption{The illustration of (a) Domain Negotiation, (b) Alternate Training, (c) MAML, and (d) Reptile.}
    \label{fig:dn}
\end{figure*}

\subsection{Theoretical Analysis}\label{sec:dn_analysis}

In this section, we first provide theoretical analyses to explain why DN could mitigate the domain conflict problem. Extending the analysis, we also provide explanations for DR.

To mitigate domain conflict, DN tries to maximize the inner-products of gradients between domains, which can be formulated as:
\begin{equation}
    \setlength\abovedisplayskip{1pt}
    \setlength\belowdisplayskip{1pt}
    \mathcal{O}_C = \max\sum_i^n\sum_j^n \langle g_i, g_j\rangle.\label{eq:dn_max_inner_product}
\end{equation}
Clearly, directly optimizing the $\mathcal{O}_C$ requires a $O(n^2)$ complexity. In DN, we first sequentially perform training on each domain in the inner loop. This provides a sequence of loss $L(\widetilde{\Theta}_{i}, T^i)$ that can be simplified as $L_i(\widetilde{\Theta}_i)$. We could define the following notations to facilitate analysis:
{
\begin{align}
    g_i                  & = L'_i(\widetilde{\Theta}_i) & & \text{(gradients from domain $i$),}                  \\
    \overline{g}_i       & = L'_i(\widetilde{\Theta}_1)                      & & \text{(gradients at initial point $\widetilde{\Theta}_{1}$),} \\
    \overline{H}_i       & = L^{''}_i(\widetilde{\Theta}_1)                  & & \text{(Hessian Matrix at initial point $\widetilde{\Theta}_{1}$),}                   \\
    \widetilde{\Theta}_i & = \widetilde{\Theta}_1 -\alpha\sum_{j=1}^{i-1}g_j & & \text{(sequence of gradient descent).}
\end{align}
}
We can perform the Taylor expansion on the $g_i$ when $\alpha$ is small enough, which is formulated as:
{
\begin{align}
    g_i & = L'_i(\widetilde{\Theta}_1) + L^{''}(\widetilde{\Theta}_1)(\widetilde{\Theta}_i - \widetilde{\Theta}_1) + O(\alpha^2),\label{eq:taylor}                        \\
        & = \overline{g}_i + \overline{H}_i(\widetilde{\Theta}_i - \widetilde{\Theta}_1) + O(\alpha^2),                                     \\
        & = \overline{g}_i - \alpha\overline{H}_i\sum_{j=1}^{i-1}g_j + O(\alpha^2),                                                                             \\
        & = \overline{g}_i - \alpha\overline{H}_i\sum_{j=1}^{i-1}\overline{g}_j + O(\alpha^2)\ \ \text{(using $g_j=\overline{g}_j + O(\alpha))$}.\label{eq:g_i}
\end{align}
}
Then, the gradients $\widetilde{\Theta}_{n+1} - \Theta$ for outer loop can be formulated as:
{
\setlength\abovedisplayskip{1pt}
\setlength\belowdisplayskip{1pt}
\begin{equation}
    \setlength\abovedisplayskip{1pt}
    \setlength\belowdisplayskip{1pt}
    \begin{split}
        -(\widetilde{\Theta}_{n+1} - \Theta)/\alpha = \sum_{i=1}^n g_j
        = \sum_{i=1}^n \overline{g}_i - \alpha\sum_{i=1}^{n}\sum_{j=1}^{i-1}\overline{H}_i\overline{g}_j\\ + O(\alpha^2).
    \end{split}
\label{eq:dn_grad}
\end{equation}
}
In equation \ref{eq:dn_grad}, the first term $\sum_{i=1}^n \overline{g}_i$ serves to jointly minimize loss on each domain. This makes sure the convergence of DN.
The second term $\sum_{i=1}^{n}\sum_{j=1}^{i-1}\overline{H}_i\overline{g}_j$, which is more interesting, serves to maximize the inner-products of gradients between domains. Thus, we denote the expectation of $\overline{H}_i\overline{g}_j$ as the \textit{InnerGrad}.
Since the sequence of domains is shuffled at every epoch in the inner loop, the InnerGrad can be formulated as:
{
    \begin{align}
        \text{InnerGrad}=\mathbb{E}(\overline{H}_i\overline{g}_j) & = \mathbb{E}(\overline{H}_j\overline{g}_i),                                                                                 \\
                                                                  & = \frac{1}{2}\mathbb{E}(\overline{H}_i\overline{g}_j + \overline{H}_j\overline{g}_i),                                       \\
                                                                  & = \frac{1}{2}\mathbb{E}\big(\frac{\partial}{\partial\Theta}\langle\overline{g}_i, \overline{g}_j\rangle\big).\label{eq:inner_grad}
    \end{align}
}

Clearly, the (-InnerGrad) is the direction that increases the inner-products of gradients between domains. Therefore, the gradient shown in equation \ref{eq:dn_grad} can not only minimize $\mathcal{O}_M$ for multi-domain recommendation, but also maximize $\mathcal{O}_C$ for mitigating domain conflict. What is more, the overall computational complexity of DN is $O(n)$, which makes it more suitable for large-scale applications.

Noticeably, as shown in Figure \ref{fig:dn} (a) and (b), the conventional Alternate Training \cite{long2021multi} directly optimizes $\Theta$ on different domains one-by-one, whereas DN consists of two training loops. The reason is that the ability of equation \ref{eq:dn_grad} to maximize gradients' inner production only holds when applied to the initial point of $\Theta$. Thus, Alternate Training without the outer loop cannot address the domain conflict issue. Besides, in the Alternate Training, $\Theta$ is easily dominant by data-richness domains \cite{ke2022domain} and cannot generalize for all domains.

Similar conclusion about maximizing the gradients' inner production has also been used by meta-learning frameworks (i.e., MAML \cite{finn2017model} and Reptile \cite{nichol2018first}). However, as shown in Figure \ref{fig:dn} (c) and (d), MAML and Reptile maximize the gradients' inner production within the same domain, which only improves its single-domain generalizability, but is unable to mitigate conflict across domains. In DN, we successfully maximize gradients' inner production across domains, which is a key contribution to mitigating domain conflict.

The theoretical analysis of DR can be extended from equation \ref{eq:g_i} and \ref{eq:dn_grad}. For each $D^j$, we first optimize $\widetilde{\theta^i}$ on $D^j$, then update on target domain $D^i$ for regularization. The gradients $\widetilde{\theta^i} - \theta^i$ for updating specific parameters can be formulated as:
\begin{equation}
    -(\widetilde{\theta^i} - \theta^i)/\alpha = g_j + g_i = \overline{g}_j + \overline{g}_i - \alpha\overline{H}_i\overline{g}_j. \label{eq:dr_grad}
\end{equation}
Because the update sequence of $D^j$ and $D^i$ is fixed, the conclusion in equation \ref{eq:inner_grad} can not be applied to $\overline{H}_i\overline{g}_j$ here. The $\overline{H}_i$ denotes the Hessian matrix of target domain, thus the $\overline{H}_i\overline{g}_j$ in DR regularizes the gradients $\overline{g}_j$ of $D^j$ to best serve the optimization of target domain $D^i$. In this way, DR can easily adopt data from other domains to improve performance on the target domain $D^i$.

\subsection{The \ourmethod Algorithm}\label{sec:mamdr}
The DN and DR could be integrated into a unified framework, i.e., \ourmethod. The overall process of \ourmethod is illustrated in Algorithm \ref{alg:mamdr}. Given $n$ different domains and arbitrary model structures with parameters $\Theta$, we copy $\Theta$ into the shared parameters $\theta^S$ and specific parameters $\{\theta^1,\cdots,\theta^n\}$. In each iteration, we first update shared parameters $\theta^S$ using Domain Negotiation (line \ref{line:dn}) to mitigate domain conflict. Then, for each specific parameter, we update $\theta^i$ using Domain Regularization (line \ref{line:dr}) to improve generalizability. The overall complexity of \ourmethod is $O\big((k+1)n\big)$. From Algorithm \ref{alg:mamdr}, we can see that our \ourmethod is agonist to model structure and can be applied to any MDR method to meet varied circumstances.

\begin{algorithm}[t]
    \caption{\ourmethod}\label{alg:mamdr}
    \KwIn {$n$ different domains $\mathcal{D}$, shared parameters $\theta^S$, domain-specific parameters $\{\theta^1,\cdots,\theta^n\}$, learning rate $\alpha,\beta,\gamma$, sample size $k$, maximum training epoch $N$.}
    \KwOut{$\Theta=\big\{\theta^S,\{\theta^1,\cdots,\theta^n\}\big\}$}
    \For{$epoch=1,\cdots,N$}{
        Update $\theta^S$ using Domain Negotiation (Algorithm \ref{alg:dn})\label{line:dn}\;
        \For{$i=1,\cdots,n$}{
            Update $\theta^i$ using Domain Regularization (Algorithm \ref{alg:dr})\label{line:dr}\;
        }
    }
    \Return{$\Theta=\big\{\theta^S,\{\theta^1,\cdots,\theta^n\}\big\}$}
\end{algorithm}

\subsection{Large-scale Implementation}\label{sec:implementation}

To support large-scale applications, we adopt the PS-Worker architecture \cite{li2014scaling} for distribute training. PS-Worker is a commonly used data-parallel method for scaling model training on multiple machines, which contains two parts of machine: parameter servers and workers. Parameter servers store model parameters, and workers calculate the gradients. 

\begin{figure}[]
    \centering
    \includegraphics[trim=0cm 0cm 0cm 0,clip,width=0.7\columnwidth]{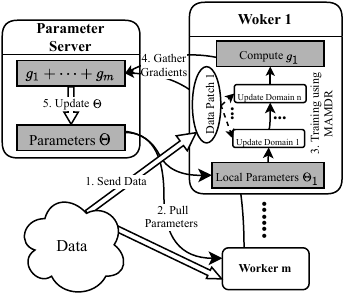}
    \caption{Implementation of \ourmethod in PS-Worker architecture.}
    \label{fig:psworker}
\end{figure}

The overall architecture is illustrated in Figure \ref{fig:psworker}. (1) we first distribute the training data into $m$ worker machines. (2) each worker obtains the parameters from the parameter server and stores as the local parameter. (3) the \ourmethod algorithm is implemented in each worker to update the local parameters and compute the local gradient. (4) workers send their local gradients to the parameter server. (5) the parameter server synthesizes the local gradients and updates the model parameters. In this way, we can implement \ourmethod in our applications to deal with billions of data.

\begin{figure}[]
    \centering
    \includegraphics[trim=0cm 0cm 0cm 0,clip,width=0.7\columnwidth]{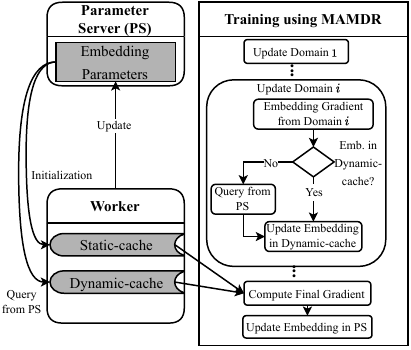}
    \caption{The illustration of Embedding PS-Worker cache.}
    \label{fig:cache}
\end{figure}

\noindent\textbf{Embedding PS-Worker Cache.}
During training, we need to update the parameters of user/item embeddings. The embedding parameters are large and sparse, which costs a lot of time to synchronize with the parameter server. Besides, the embeddings are actively updated across different workers, which could lead to the inconsistency of embedding parameters due to the asynchronous update. To reduce the synchronization overhead and alleviate inconsistency, we propose an embedding PS-Worker cache mechanism to cache the embedding parameters in worker machines and speed up parameters' synchronization. The process is illustrated in Figure \ref{fig:cache}. 

Specifically, we define a \textit{dynamic-cache} and a \textit{static-cache} for each worker, which stores the embedding parameters. The static-cache is initialized by obtaining the parameters from the PS and remained unchanged during the training process of \ourmethod, while the dynamic-cache is updated in the inner loop of \ourmethod. In inner loop, we compute embedding gradients from each domain. For embedding parameters to be updated, we first check whether the embedding parameters are in the dynamic-cache. If yes, we directly update the embedding in dynamic-cache. If not, we query the latest embedding from the PS, then we update the embedding and cache it in dynamic-cache. After the inner loop, we use the parameters in dynamic-cache and static-cache to compute the final gradients in outer loop and update the parameters in PS using Equation \ref{eq:outer}. Last, we clear both the static-cache and dynamic-cache for next epoch.
In this way, we can not only reduce the synchronization overhead but also alleviate inconsistency by querying the latest embedding from the parameter server on demand.

\noindent\textbf{Parameters Update}
For other model parameters, the worker obtains the parameters from the PS at the beginning of each epoch and stores them in the static-cache as $\Theta$. In inner loop, we use dedicated optimizer (e.g., SGD) to optimize parameters and store the updated parameter $\Theta_i$ in dynamic-cache. In outer loop, we use parameters stored in dynamic-cache and static-cache to compute gradients $\Theta_{n+1} - \Theta$. Then, we empty the caches and send the gradients back to PS. The parameters in PS are updated using Equation \ref{eq:outer} with another optimizer (e.g., Adagrad). In this way, both the learning rate and optimizer in the inner and outer loop can be independently specified, which is more suitable for the model agnostic setting.

\section{Experiment}
In experiments, we conduct CTR prediction (one of the typical recommendation tasks) to evaluate the performance of the proposed \ourmethod. Code and data used in experiments are available at: \url{https://github.com/RManLuo/MAMDR}.

\subsection{Dataset}
\noindent\textbf{MDR benchmark datasets.}
In experiments, we first construct several MDR benchmark datasets based on public real-world data (i.e., Amazon \cite{he2016ups} and Taobao \cite{du2019sequential}). The Amazon dataset is a large-scale dataset containing product reviews collected from Amazon.com. We split domains by the category of products such as ``Musical Instruments'' and ``Video Games''. For each domain, we keep the existing user-product reviews as positive samples. The Taobao dataset is from the click log of Cloud Theme in the Taobao app. It has already split domains by the theme of purchase, e.g.,``what to take when traveling'', ``how to dress up yourself for a party'', and ``things to prepare when a baby is coming''. Each domain contains a set of users and items, thus we treat the positive samples as the items clicked by users. As shown in Figure \ref{fig:framework}, our MDR system maintains a global features storage. Thus, we unify the users/items' features from all domains together. For Amazon datasets, due to the lack of features, we randomly initialize the embeddings as features and optimize them during training. For Taobao datasets, the features are built based on users' attributes and historical interactions in Taobao by GraphSage \cite{hamilton2017inductive}. We fixed these features during training.

In MDR benchmark datasets, we want to simulate real-world situations based on the challenges we found in our industry applications (i.e., domain conflict and imbalanced data distribution). Thus, for the Amazon dataset, we first select 6 domains that have relatively abundant interactions to simulate the normal data distribution and construct the Amazon-6. Then, 7 domains with fewer interactions are added to simulate the data sparsity situation and form the Amazon-13. As for the Taobao dataset, we randomly select 10, 20, and 30 domains from the dataset to form three sub-datasets: Taobao-10, Taobao-20, and Taobao-30. In this way, we try to approximate the real data distribution in the industry scenario. For each dataset, we randomly select items that haven't been reviewed or clicked by users as negative samples. The number of negative samples is selected by the given CTR Ratio for each domain. The CTR ratio is defined as:
\begin{equation}
    \text{CTR Ratio} = \frac{\text{\#positive samples}}{\text{\#negative samples}}.
\end{equation}
We randomly assign CTR Ratio for each domain ranging from $[0.2,0.5]$ to simulate the domain distinction.

\begin{table}[]
    \centering
    \Huge
    \caption{Overall statistic of datasets.}
    \label{tab:dataset}
    \resizebox{1\linewidth}{!}{%
        \begin{tabular}{cccccccc}
            \toprule
            Dataset   & \#Domain & \#User    & \#Item    & \#Train    & \#Val     & \#Test    & Sample / Domain \\
            \midrule
            Amazon-6  & 6         & 445,789    & 172,653    & 9,968,333   & 3,372,666  & 3,585,877  & 2,821,146       \\
            Amazon-13 & 13        & 502,222    & 215,403    & 11,999,607  & 4,100,756  & 4,339,523  & 1,572,299       \\\midrule
            Taobao-10 & 10        & 23,778     & 6,932      & 92,137      & 37,645     & 43,502     & 17,328          \\
            Taobao-20 & 20        & 58,190     & 16,319     & 243,592     & 96,591     & 106,500    & 22,334          \\
            Taobao-30 & 30        & 99,143     & 29,945     & 394,805     & 151,369    & 179,252    & 24,180          \\\midrule
            Taobao-online  & 69,102    & 84,307,785 & 16,385,662 & 420,097,203 & 23,340,352 & 46,415,298 & 7,088           \\
            \bottomrule
        \end{tabular}
    }
\end{table}

\begin{table*}[]
    \centering
    \Huge
    \caption{Statistics of Amazon-6 dataset.}
    \label{tab:amazon6}
    \resizebox{0.6\linewidth}{!}{%
        \begin{tabular}{@{}ccccccc@{}}
            \toprule
            Domain     & Musical Instruments & Office Products & Patio Lawn and Garden & Prime Pantry & Toys and Games & Video Games \\ \midrule
            \#Samples & 1,204,340             & 3,921,259         & 3,025,218               & 694,758       & 5,382,501        & 2,698,800     \\
            Percentage & 7.11\%              & 23.17\%         & 17.87\%               & 4.10\%       & 31.80\%        & 15.94\%     \\
            CTR Ratio  & 0.22                & 0.23            & 0.32                  & 0.23         & 0.47           & 0.21        \\ \bottomrule
        \end{tabular}
    }
\end{table*}

\begin{table*}[]
    \centering
    \caption{Statistics of Amazon-13 dataset.}
    \label{tab:amazon13}
    \resizebox{.7\linewidth}{!}{%
        \begin{tabular}{@{}cccccccc@{}}
            \toprule
            Domain     & Arts Crafts and Sewing & Digital Music & Gift Cards & Industrial and Scientific & Luxury Beauty & Magazine Subscriptions   \\ \midrule
            \#Samples & 2,419,005                & 770,132        & 11,951      & 380,386                    & 87,360         & 13,103                                \\
            Percentage & 11.86\%                & 3.78\%        & 0.06\%     & 1.86\%                    & 0.43\%        & 0.06\%                               \\
            CTR Ratio  & 0.22                   & 0.23          & 0.32       & 0.23                      & 0.47          & 0.21                  \\ \bottomrule
        \end{tabular}
    }\\[1pt]
    \resizebox{.7\linewidth}{!}{%
    \begin{tabular}{@{}cccccccccccccc@{}}
        \toprule
        Domain     & Musical Instruments & Office Products & Patio Lawn and Garden & Prime Pantry & Software & Toys and Games & Video Games \\ \midrule
        \#Samples & 814,928 & 3,178,096         & 2,317,603               & 655,970       & 11,022    & 7,541,261        & 459,646      \\
        Percentage  & 3.99\% & 15.58\%         & 11.36\%               & 3.22\%       & 0.05\%   & 36.97\%        & 10.78\%      \\
        CTR Ratio               & 0.36   & 0.30            & 0.46                  & 0.25         & 0.30     & 0.30           & 0.27        \\ \bottomrule
    \end{tabular}
    }
\end{table*}

\begin{table*}[]
    \centering

    \caption{Statistics of Taobao-10/20/30 dataset. The first 10 and 20 domains are used for Taobao-10 and Taobao-20, respectively.}
    \label{tab:taobao}
    \resizebox{.7\linewidth}{!}{%
        \begin{tabular}{@{}cccccccccccccccc@{}}
            \toprule
            Domain     & D1     & D2     & D3     & D4     & D5     & D6     & D7     & D8     & D9     & D10    & D11    & D12    & D13    & D14     & D15    \\ \midrule
            \#Sample  & 13,255  & 7,007   & 20,130  & 62,460  & 11,563  & 7,190   & 4,185   & 24,052  & 5,582   & 17,860  & 29,302  & 6,472   & 8,873   & 125,586  & 15,560  \\
            Percentage & 1.82\% & 0.96\% & 2.77\% & 8.60\% & 1.59\% & 0.99\% & 0.58\% & 3.31\% & 0.77\% & 2.46\% & 4.03\% & 0.89\% & 1.22\% & 17.29\% & 2.14\% \\
            CTR Ratio  & 0.22   & 0.23   & 0.32   & 0.23   & 0.47   & 0.21   & 0.36   & 0.30   & 0.46   & 0.25   & 0.30   & 0.30   & 0.27   & 0.20    & 0.33  \\ \bottomrule
        \end{tabular}
    }\\[2pt]
    \resizebox{.7\linewidth}{!}{%
        \begin{tabular}{@{}cccccccccccccccc@{}}
            \toprule
            Domain     & D16    & D17    & D18    & D19    & D20    & D21    & D22    & D23    & D24    & D25    & D26    & D27    & D28    & D29    & D30    \\ \midrule
            \#Sample  & 5,458   & 14,095  & 53,910  & 12,102  & 2,936   & 4,710   & 29,256  & 41,609  & 7,354   & 68,119  & 5,308   & 24,918  & 38,919  & 24,297  & 34,253  \\
            Percentage & 0.75\% & 1.94\% & 7.42\% & 1.67\% & 0.40\% & 0.65\% & 4.03\% & 5.73\% & 1.01\% & 9.38\% & 0.73\% & 3.43\% & 5.36\% & 3.35\% & 4.72\% \\
            CTR Ratio & 0.23   & 0.38   & 0.22   & 0.29   & 0.33   & 0.47   & 0.23   & 0.24   & 0.44   & 0.21   & 0.47   & 0.37   & 0.28   & 0.45   & 0.43   \\ \bottomrule
        \end{tabular}
    }
\end{table*}

\noindent\textbf{Industry dataset.}
In addition to the MDR benchmark datasets, we also conduct experiments on an industry dataset (i.e., Taobao-online). This dataset contains 69,102 different domains and 489,852,853 samples, which is collected from the Taobao app, one of the largest online shopping applications in China. Domains in this large-scale industry dataset are more complicated and share diverse relatedness, which can better reflect the real-world scenario. The statistics of industry dataset and MDR benchmark datasets are summarized in Table \ref{tab:dataset}. The click distribution and statistics of datasets are illustrated in Table \ref{tab:amazon6}, \ref{tab:amazon13}, and \ref{tab:taobao}. Domains with smaller CTR ratios are sparser.

\subsection{Baselines}
We select several state-of-the-art methods in CTR prediction as baselines, which can be roughly grouped into two categories: Single-Domain methods and Multi-Task (Domain) methods.

\noindent\textbf{Single-Domain Method.} This group of methods is originally proposed for single domain recommendation. Thus, they do not consider the domain distinction information.
\begin{itemize}
    \item MLP: Multi-layer perceptron (MLP) is the simplest neural network model composed of multiple fully connected layers.
    \item WDL \cite{cheng2016wide}: WDL is a widely used recommendation model in the industry. It contains a wide liner network and a deep neural network to simultaneously capture the cross-product features as well as the nonlinear features.
    \item NeurFM \cite{he2017neural}: NeurFM proposes a bi-interaction pooling layer to learn feature interaction between embedding vectors. Then, it integrates the results with the logit output of a MLP layer.
    \item AutoInt \cite{song2019autoint}: AutoInt proposes the attention-based interaction layer to automatically identify meaningful high-order features.
    \item DeepFM \cite{guo2017deepfm}: integrate the factorization machine with the deep neural network to improve the recommendation results.
\end{itemize}

\noindent\textbf{Multi-Task (Domain) Method.} This group of methods contains the multi-task and multi-domain methods. As discussed in the Introduction, the multi-task methods can be directly applied to MDR by treating each domain as a separate task.
\begin{itemize}
    \item Shared-Bottom \cite{ruder2017overview}: Shared-Bottom is a multi-task method that consists of shared-bottom networks and several domain-specific tower networks. Each domain has its specific tower network while sharing the same bottom network.
    \item MMOE \cite{ma2018modeling}: MMOE adopts the  Mixture-of-Experts (MoE) structure by sharing the expert modules across all domains, while having a gating network trained for each domain.
    \item PLE \cite{tang2020progressive}: PLE separates shared components and task-specific components explicitly and adopts a progressive mechanism to extract features gradually.
    \item Star \cite{sheng2021one}: is the state-of-the-art MDR method. It splits the parameters into shared and specific parts. Meanwhile, it proposes a Partitioned Normalization for distinct domain statistics.
\end{itemize}

To compare with other model agnostic learning frameworks, we select several representative learning frameworks, which can be roughly grouped into three categories: traditional learning frameworks, multi-task leaning methods, and meta-learning frameworks. 

\noindent\textbf{Traditional Learning Frameworks.}
\begin{itemize}
    \item Alternate: Alternate learning is a conventional learning framework that trains on multiple domains alternately. In this way, it enables the model to learn shared features and improve performance.
    \item Alternate + Finetune: On the top of the model learned by Alternate learning, we finetune the model on each domain to obtain several domain-specific models. Each specific model can capture the domain distinction information.
\end{itemize}

\noindent\textbf{Multi-Task Learning Frameworks.}
\begin{itemize}
    \item Weighted Loss \cite{kendall2018multi}: Weighted Loss is a multi-task learning framework that assigns weight to the loss of each task. Meanwhile, the weight is automatically optimized during the training to balance between different losses.
    \item PCGrad \cite{yu2020gradient}: PCGrad is a powerful multi-task learning framework. By projecting the gradients of each task into the normal plane of others, it successfully avoids the conflicting problem.
\end{itemize}

\noindent\textbf{Meta-Learning Frameworks.}
\begin{itemize}
    \item MAML \cite{finn2017model}: MAML aims to train parameters through various tasks and acquires parameters that can quickly adapt to new tasks. We treat each domain as the task and split the training data into the support and query sets used for MAML.
    \item Reptile \cite{nichol2018first}: Reptile is a first-order meta-learning framework, which trains parameters by rapidly sampling the tasks. It maximizes the inner-gradients within the task and leads parameters quickly to generalize to new tasks.
    \item MLDG \cite{li2018learning}: MLDG proposes a novel meta-learning framework for domain generalization. Its meta-optimization function improves the performance on both train and test domains.
\end{itemize}

\begin{table*}[]
    \centering
    \caption{Comparison with multi-domain recommendation methods under average AUC and average RANK metrics.}
    \label{tab:cmp}
    \begin{tabular}{@{}c|cc|cc|cc|cc|cc@{}}
    \toprule
    \multirow{2}{*}{Method} & \multicolumn{2}{c|}{Amazon-6}  & \multicolumn{2}{c|}{Amazon-13} & \multicolumn{2}{c|}{Taobao-10} & \multicolumn{2}{c|}{Taobao-20} & \multicolumn{2}{c}{Taobao-30}  \\ \cmidrule(l){2-11} 
                            & AUC             & RANK         & AUC             & RANK         & AUC             & RANK         & AUC             & RANK         & AUC             & RANK         \\ \midrule
    MLP                     & 0.7464          & 9.0          & 0.7016          & 8.6          & 0.7022          & 11.3         & 0.7255          & 9.9          & 0.7416          & 10.7         \\
    WDL                     & 0.7449          & 9.0          & 0.7026          & 7.9          & 0.7154          & 8.9          & 0.7235          & 10.6         & 0.7559          & 8.4          \\
    NeurFM                  & 0.6505          & 10.7         & 0.6152          & 10.2         & 0.7374          & 4.1          & 0.7461          & 6.4          & 0.7673          & 6.1          \\
    AutoInt                 & 0.7531          & 8.2          & 0.7214          & 6.4          & 0.7302          & 5.8          & 0.7471          & 6.3          & 0.7623          & 6.5          \\
    DeepFM                  & 0.7333          & 10.0         & 0.6976          & 8.5          & 0.7271          & 6.6          & 0.7347          & 8.8          & 0.7484          & 9.4          \\\midrule
    Shared-bottom           & 0.7794          & 3.0          & 0.7088          & 5.0          & 0.7197          & 7.7          & 0.7572          & 4.3          & 0.7714          & 6.1          \\
    MMOE                    & 0.7816          & 2.7          & 0.7381          & 4.2          & 0.7250          & 5.9          & 0.7494          & 6.0          & 0.7717          & 4.2          \\
    PLE                     & 0.7801          & 3.5          & 0.7114          & 6.3          & 0.7287          & 5.3          & 0.7603          & 3.3          & 0.7725          & 4.0          \\
    Star                    & 0.7719          & 5.8          & 0.7209          & 7.1          & 0.7202          & 8.0          & 0.7324          & 8.9          & 0.7483          & 9.4          \\\midrule
    MLP+\ourmethod       & \textbf{0.7957} & \textbf{2.5} & \textbf{0.7577} & \textbf{3.5} & \textbf{0.7445} & \textbf{2.7} & \textbf{0.7613} & \textbf{3.2} & \textbf{0.7750} & \textbf{3.1} \\ \bottomrule
    \end{tabular}
\end{table*}
\subsection{Implementation Details}\label{app:parameters}
All the models in single-domain and multi-domain methods, except the Star, are implemented by DeepCTR \cite{shen2017deepctr}, which is an open-source deep-learning based CTR package. Star is implemented by us according to the original paper. We implement all the learning frameworks in Tensorflow.

To make a fair comparison, we try to set similar parameters for all the models. For all single-domain methods, the hidden layers are set to $[256,128,64]$; for AutoInt, its attention head number is set to $4$; for Shared-bottom, its shared network is set to $[512,256,128]$ and its tower network is set to $[64]$; for MMOE, its expert networks are set to $[512,256,128]$, its tower network and gating network are set to $[64]$, and its expert number is set to $2$; for PLE, its shared network is set to $[64]$, its tower network is set to $[256]$, its gating network is set to $[64]$, its shared and specific experts are set to $2$ and $10$, respectively; for Star, both the shared and specific networks are set as $[256,128,64]$. For all models, the dropout rate are set to 0.5, the embedding size is set to 128, and the learning rate is set to 0.001. For our \ourmethod, the inner learning rate is set to 0.001, and the outer learning rate is set to 0.1; the sample number of DR is set to $[3,5,5,5,5]$ for each dataset respectively. We use Adam as the optimizer and Binary Cross Entropy as the recommendation loss function.

In the industry dataset, we use SGD for inner loop with learning rate set to 0.1, and Adagrad \cite{duchi2011adaptive} for the outer loop with a dynamical learning rate ranging from 0.1 to 1. The batch size is set to 1024. The feature size is set to 1700. We use 40 parameter servers, each with 30 CPUs and 50GB memory, and 400 workers, each with 20 CPUs and 100GB memory. Adding a new domain will introduce 20M parameters to the model.

\subsection{Results in MDR Benchmark datasets}\label{sec:performance}
We first conduct CTR prediction to evaluate the performance of the proposed \ourmethod on MDR benchmark datasets. The area under the ROC curve (AUC) is the common metric used to evaluate the performance of CTR prediction. Thus, average AUC of all domains and average performance RANK among baselines of all domains are selected as the final metrics.
%
Since our \ourmethod is agnostic to model structure, we just use the simplest multi-layer perceptron (MLP) with three fully connected layers as the base model structure. These baselines are alternately trained using data from all domains.

The comparison results are shown in Table \ref{tab:cmp}, where the best results are highlighted with bold type. From results, we can see that \ourmethod (DN + DR) greatly improves the AUC of MLP and outperforms other baselines in RANK throughout all datasets.
Compared to Amazon-6, with the number of domains increasing, both the performance of single-domain and multi-domain methods deteriorates in Amazon-13.
This is because 7 sparse domains are introduced into Amazon-13, which makes the specific parameters overfitting.
 \ourmethod (DN+DR) takes advantage of the information from other domains to mitigate overfitting, which boosts the improvement of MLP by 6.6\% and 8.0\% in Amazon-6 and Amazon-13, respectively.

In Taobao dataset, the performance of each model improves with domains' numbers increasing. The possible reason is that training samples of each domain are sparser in Taobao dataset as shown in the Table \ref{tab:dataset}. Therefore, more domains introduce more training samples and improve the overall performance. This also indicates the importance of shared information for multi-domain recommendations. 
Although more domains could facilitate the performance, it also increases the possibility of domain conflict. \ourmethod (DN+DR) not only alleviates the domain conflict (DN) but also improves generalizability on sparse domains (DR), which promotes the performance of MLP to the best place among all Taobao datasets.

\begin{table*}[]
    \centering
    \caption{Ablation study of DN and DR.}
    \label{tab:abl}
    \begin{tabular}{@{}c|cc|cc|cc|cc|cc@{}}
    \toprule
    \multirow{2}{*}{Method} & \multicolumn{2}{c|}{Amazon-6}  & \multicolumn{2}{c|}{Amazon-13} & \multicolumn{2}{c|}{Taobao-10} & \multicolumn{2}{c|}{Taobao-20} & \multicolumn{2}{c}{Taobao-30}  \\ \cmidrule(l){2-11} 
                            & AUC             & RANK         & AUC             & RANK         & AUC             & RANK         & AUC             & RANK         & AUC             & RANK         \\ \midrule
    MLP+\ourmethod (DN+DR)       & \textbf{0.7957} & \textbf{2.5} & \textbf{0.7577} & \textbf{3.5} & \textbf{0.7445} & \textbf{2.7} & \textbf{0.7613} & \textbf{3.2} & \textbf{0.7750} & \textbf{3.1} \\ 
    $w/o$ DN                 & 0.7822          & 5.2          & 0.7507          & 4.8          & 0.7407          & 4.0          & 0.7596          & 3.9          & 0.7501          & 3.8          \\ 
    $w/o$ DR                  & 0.7678          & 8.5          & 0.7331          & 5.4          & 0.7204          & 7.7          & 0.7501          & 6.5          & 0.7619          & 6.3          \\
    $w/o$ DN+DR & 0.7464          & 9.0          & 0.7016          & 8.6          & 0.7022          & 11.3         & 0.7255          & 9.9          & 0.7416          & 10.7         \\
    \bottomrule
    \end{tabular}
\end{table*}

\begin{table*}[]
    \centering
    \caption{Results of each domain on Amazon-6.}
    \label{tab:domain}    
    \resizebox{.85\linewidth}{!}{%
    \begin{tabular}{@{}c|cccccc@{}}
    \toprule
    Method            & Musical Instruments & Office Products & Patio Lawn and Garden & Prime Pantry    & Toys and Games  & Video Games     \\ \midrule
    MLP+\ourmethod (DN+DR) & \textbf{0.7753}     & \textbf{0.8116} & \textbf{0.7579}       & \textbf{0.7579} & \textbf{0.8108} & \textbf{0.8394} \\
    $w/o$ DN          & 0.7617              & 0.7940          & 0.7583                & 0.7498          & 0.7918          & 0.8371          \\
    $w/o$ DR          & 0.7544              & 0.7840          & 0.7533                & 0.7140          & 0.7602          & 0.7763          \\
    $w/o$ DN+DR               & 0.7223              & 0.7257          & 0.7509                & 0.7171          & 0.7423          & 0.7804          \\ \bottomrule
    \end{tabular}%
    }

\end{table*}

Last, even some MDR models have complex structures (e.g., NeurFM, AutoInt, MMOE, and PLE), their performance cannot outperform a simple MLP optimized under proposed \ourmethod. What is more, their performances are diverse from different datasets. This indicates that the existing models' structure is not suitable for all circumstances. In contrast, \ourmethod has no restriction on model structure and could easily fit any datasets without burdensome hyper-parameters turning.

\subsection{Ablation Study}\label{sec:abl}
To analyze the effectiveness of DN and DR, we first conduct the ablation study. We select MLP as the base model and the results of different model variances are shown in Table \ref{tab:abl}. By removing the DN, the shared parameters could suffer from domain conflict and impairs the performance. This is more likely to happen with domain number increases (e.g., Taobao-30). Without DR, the specific parameters are inclined to overfit as shown in Amazon-13 which has 7 sparse domains. Last, we can see that both the DN and DR are able to improve the performance of MLP. Thus, we should combine them and use \ourmethod (DN+DR) to achieve the best performance and generalizability. 

We further show specific result of each domain on Amazon-6 in Table \ref{tab:domain}. From results, we can see that MLP+\ourmethod achieves the best performance throughout all domains. Both removing the DN and DR would lead to performance drops. Noticeably, the performance of the domain ``Prime Pantry'', which has fewer samples, abates significantly (5.79\%) when removing the DR. This also demonstrates the effectiveness of DR on sparse domains.

\begin{table*}
    \centering
    \caption{Results on the industry dataset under average AUC metric.}
    \label{tab:online}
    \resizebox{0.6\linewidth}{!}{%
        \begin{tabular}{@{}c|ccccccc@{}}
            \toprule
            Methods & RAW & MMOE   & CGC    & PLE   & RAW+Separate  & RAW+DN & RAW+\ourmethod       \\ \midrule
            AUC          & 0.7503        & 0.7497 & 0.7489 & 0.7513  & 0.7460 & 0.7559 & \textbf{0.7700}
            \\ \bottomrule
        \end{tabular}
    }
\end{table*}
\begin{table*}
    \centering
    \caption{Results on top 10 largest domains of industry dataset under AUC metric.}
    \label{tab:detailed}
    \resizebox{0.65\linewidth}{!}{%
        \begin{tabular}{@{}c|cccccccccc@{}}
            \toprule                                                                                
            Methods       & Top 1              & Top 2              & Top 3              & Top 4              & Top 5              & Top 6              & Top 7              & Top 8              & Top 9              & Top 10             \\ \midrule
            RAW & 0.8202          & 0.7635          & 0.8439          & 0.7295          & 0.6962          & 0.7417          & 0.6661          & 0.7524          & 0.7540          & 0.6912          \\
            MMOE          & 0.8166          & 0.7597          & 0.8288          & 0.7694          & 0.6945          & 0.7453          & 0.6677          & 0.7315          & 0.7478          & 0.6941          \\
            CGC           & 0.8172          & 0.7640          & 0.8307          & 0.7747          & 0.7215          & 0.7392          & 0.6726          & 0.7444          & 0.7357          & 0.7019          \\
            PLE           & 0.8158          & 0.7643          & 0.8261          & 0.7768          & 0.7327          & 0.7284          & 0.6793          & 0.7410          & 0.7472          & 0.7038          \\\midrule
            RAW+Separate  & 0.8127          & 0.7635          & 0.8285          & 0.7569          & 0.6896          & 0.7367          & 0.6701          & 0.7370          & 0.7283          & 0.6947          \\
            RAW+DN        & 0.8173          & 0.7655          & 0.8397          & 0.7643          & 0.7188          & 0.7344          & 0.6664          & 0.7523          & 0.7505          & 0.7021          \\
            RAW+\ourmethod     & \textbf{0.8226} & \textbf{0.7704} & \textbf{0.8469} & \textbf{0.8090} & \textbf{0.7391} & \textbf{0.7648} & \textbf{0.6965} & \textbf{0.7666} & \textbf{0.7689} & \textbf{0.7150} \\ \bottomrule
        \end{tabular}%
    }
\end{table*}

\subsection{Results in Industry Dataset}\label{sec:online}
To evaluate the performance of \ourmethod on real-world scenarios, We have implemented \ourmethod in Taobao and conducted experiments on industry datasets.
We apply \ourmethod to our existing recommender model used in online service (denoted as \textit{RAW}) and compare it with other methods (i.e., MMOE \cite{ma2018modeling}, CGC \cite{tang2020progressive}, and PLE \cite{tang2020progressive}) in industry dataset. All the baselines are trained using alternate training.

We first show the average AUC of 69,102 domains in Table \ref{tab:online}, where we can see that \ourmethod successfully improves the performance of existing models and reaches the best results. Besides, performance of MMOE and CGC is slightly worse. The reason is that some domains have limited samples, which could lead specific parameters to overfit on them. This is also demonstrated by separately training the model on each domain (i.e., \textit{RAW+Separate}).
Our \ourmethod can not only mitigate the domain conflict to leverage shared features but also alleviate the overfitting problem for sparse domains. 

In the Table \ref{tab:detailed}, we present the results of top 10 largest domains in online applications. From results, we can see that \ourmethod achieves the best performance among all specific domains, which shows the effectiveness of \ourmethod on data richness domains. Last, experiments on large-scale online applications also demonstrate the scalability of \ourmethod in the real-world.

\begin{table*}[]
    \centering
    \caption{Comparison with other learning frameworks under average AUC metric on Taobao-10.}
    \label{tab:method}
    \resizebox{.83\linewidth}{!}{%
        \begin{tabular}{c|cc|cc|ccc|ccc}
            \toprule
            Method        & Alternate & Alternate+Finetune & Weighted Loss & PCGrad & MAML   & Reptile & MLDG   & DN & DR     & \ourmethod (DN+DR)   \\
            \midrule
            MLP           & 0.7022    & 0.7126             & 0.7157        & 0.7254 & 0.6896 & 0.7117  & 0.7074 & 0.7204 & 0.7407 & \textbf{0.7445} \\
            WDL           & 0.7154    & 0.7040             & 0.7098        & 0.7153 & 0.6945 & 0.7212  & 0.7182 & 0.7295 & 0.7346 & \textbf{0.7376} \\
            NeurFM        & 0.7154    & 0.7465             & 0.7393        & 0.7526 & 0.7479 & 0.7579  & 0.7543 & 0.7572 & 0.7553 & \textbf{0.7609} \\
            DeepFM        & 0.7271    & 0.7280             & 0.7259        & 0.7562 & 0.7237 & 0.7402  & 0.7480 & 0.7352 & 0.7466 & \textbf{0.7581} \\
            Shared-bottom & 0.7197    & 0.7225             & 0.7171        & 0.7269 & 0.6816 & 0.7255  & 0.7195 & 0.7233 & 0.7244 & \textbf{0.7339} \\
            Star          & 0.7202    & 0.7303             & 0.7297        & 0.7221 & 0.7228 & 0.7353  & 0.7181 & 0.7328 & 0.7255 & \textbf{0.7520} \\
            \bottomrule
        \end{tabular}
    }
\end{table*}

\subsection{Learning Framework Comparison}\label{sec:method}
In this section, we will compare our \ourmethod with other model agnostic learning frameworks under different model structures. 
We conduct experiments on Taobao-10, and the results are shown in Table \ref{tab:method}.

From results, we can clearly find that \ourmethod outperforms all learning frameworks with respect to all model structures. For traditional learning frameworks, simply finetuning on each domain could improve performance for most models. But the performance of WDL slightly drops after finetuning, which may be due to the overfitting on certain domains.

Among multi-task learning frameworks, PCGrad performs better than Weighted Loss. PCGrad tries to solve domain conflict problems by projecting gradients from two domains into the non-conflict direction. But Weighted Loss only adjusts the weight of loss for different domains, which cannot fundamentally solve the domain conflict problem. In addition, Weighted Loss could give a bigger weight to the domain that is easy to train. In this way, the model could end up staying at the local optimal point of that domain rather than the global optimal point for all domains. Though the effectiveness of PCGrad, its gradient manipulation could lead the model to stay at random points. Meanwhile, the complexity of PCGrad is $O(n^2)$, which is unacceptable for large-scale MDR.

Meta-learning frameworks try to maximize the inner-product between gradients and improve generalizability. Among them, MAML achieves the worst results.
The possible reason is that MAML was originally proposed for few-shot learning on unseen domains. It splits the training samples into two subsets (query and support set), which cannot fully utilize the training sets. Reptile and MLDG, on the other hand, do not split the training samples and thus reach better results. However, they only maximize the inner-product within the same domain rather than across domains.

For our methods, we can see that DR performs better in single domain models (e.g., MLP, WDL, and NeurFM). The reason is that DR introduces specific parameters to capture domain distinction. For models containing specific parameters (e.g., Shared-bottom and Star), DN is more helpful. Because DN alleviates the domain conflict when optimizing the shared parameters. But, DR is still able to improve performance for these methods by learning from other domains. In general, we should adopt DN+DR to reach the best generalizability.


\subsection{Parameters Analysis}\label{sec:parameters}
\begin{figure}
    \centering
    \begin{minipage}[c]{0.48\columnwidth}
        \centering
    \includegraphics[trim=0 0cm 0cm 0cm,clip,width=1\linewidth]{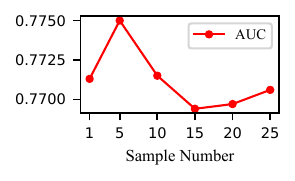}
    \end{minipage}
    \begin{minipage}[c]{0.5\columnwidth}
        \centering
    \includegraphics[trim=0.2cm 0.2cm 0.2cm 0.6cm,clip,width=1\linewidth]{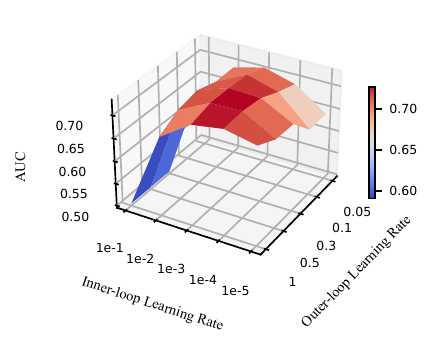}
    \end{minipage}\\
    \begin{minipage}[t]{0.48\columnwidth}	
        \caption{Results under different sample number $k$.}
        \label{fig:sample_size}
	\end{minipage}%
	\begin{minipage}[t]{0.5\columnwidth}
        \caption{Results under different learning rates.}\label{fig:lr}
	\end{minipage}%
\end{figure}

In this section, we will analyze the parameters setting of our \ourmethod. First, we analyze the number of sample domains used in Domain Regularization. Experiments are conducted on Taobao-30 with various sample number $k$. From the results shown in Figure \ref{fig:sample_size}, we can see that with the sample number increasing, the performance of the model first improves and drops at $k=5$. Because updating using too many domains will lead specific parameters deviating much from the shared parameters and impairing results. In addition, this also shows that DR would not need many domains to enhance the performance, which guarantees computational efficiency.

Second, we analyze the effect of inner-loop learning rate $\alpha$ and outer-loop learning rate $\beta$ in DN. From the results shown in Figure \ref{fig:lr}, we can see that the best performance is achieved with $\alpha=1e^{-3}$ and $\beta=[0.5,0.1]$. The reason is that according to the analysis in equation \ref{eq:g_i}, the Taylor expansion only holds when $\alpha$ is small enough. Thus, the model is barely trained when $\alpha=1e^{-1}$ or $1e^{-2}$. Besides, the results also show that using a slightly bigger $\beta$ would not impair the performance, and it can also improve training speed. Noticeably, when the outer-learning rate is set to 1, the performance drops. Because when $\beta=1$, the DN will degrade to Alternate Training in MTL, which could be affected by some data richness domains and cannot maximize the inner-gradient as DN does.

\section{Conclusion}

In this paper, we propose a novel model agnostic learning framework for multi-domain recommendation, denoted as \ourmethod. \ourmethod unifies the Domain Negotiation (DN) and Domain Regularization (DR) in the same framework to address the domain conflict and overfitting problem. We also provide a distributed implementation of \ourmethod to support large-scale applications and construct various MDR benchmark datasets, which can be used for following studies. Extensive results on MDR benchmark datasets and industry applications demonstrate both the effectiveness and generalizability of \ourmethod. Furthermore, instead of the multi-domain recommendation, the proposed DN and DR have the potential to be used for other problems such as multi-task learning, domain adaptation, and domain generalization.
\bibliographystyle{IEEEtran}
\bibliography{sections/myref.bib}

\end{document}